\documentclass[pdflatex,sn-mathphys-num]{sn-jnl}

\usepackage{graphicx}%
\usepackage{multirow}%
\usepackage{amsmath,amssymb,amsfonts}%
\usepackage{amsthm}%
\usepackage{mathrsfs}%
\usepackage[title]{appendix}%
\usepackage{xcolor}%
\usepackage{textcomp}%
\usepackage{manyfoot}%
\usepackage{booktabs}%
\usepackage{algorithm}%
\usepackage{algorithmicx}%
\usepackage{algpseudocode}%
\usepackage{braket}
\usepackage[numbers]{natbib}
\usepackage{listingsutf8}
\usepackage{yquant}
\usepackage{pgfplots}
\usepackage{subcaption}
\usepackage{csquotes}
\usepackage{adjustbox}
\definecolor{codegreen}{rgb}{0,0.6,0}
\definecolor{codegray}{rgb}{0.5,0.5,0.5}
\definecolor{codepurple}{rgb}{0.58,0,0.82}
\definecolor{backcolour}{rgb}{0.95, 0.95, 0.95}

\lstdefinestyle{codeblock}{
  backgroundcolor=\color{backcolour},
  commentstyle=\color{codegreen},
  keywordstyle=\color{blue},
  numberstyle=\tiny\color{codegray},
  stringstyle=\color{codepurple},
  basicstyle=\footnotesize,
  escapechar=\¢, 
  otherkeywords={with},
  breakatwhitespace=false,
  breaklines=true,
  captionpos=b,
  keepspaces=true,
  language=Python,
  numbers=right,
  numbersep=5pt,
  showspaces=false,
  showstringspaces=false,
  showtabs=false,
  tabsize=2,
  basicstyle=\ttfamily\footnotesize,
  inputencoding=utf8,
  upquote=true,
}
\theoremstyle{thmstyleone}%

\theoremstyle{thmstyletwo}%
\newtheorem{example}{Example}%

\theoremstyle{thmstylethree}%

\newcommand{\olivia}[1]{\textcolor{olive}{#1}}
\begin{document}

\author*[1]{\fnm{Nils} \sur{Quetschlich}}\email{nils.quetschlich@tum.de}

\author*[2]{\fnm{Olivia} \sur{Di Matteo}}\email{olivia@ece.ubc.ca}

\affil*[1]{\orgdiv{Chair for Design Automation}, \orgname{Technical University of Munich}, \orgaddress{\street{Arcisstr. 21}, \city{Munich}, \postcode{80333}, \country{Germany}}}

\affil*[2]{\orgdiv{Department of Electrical and Computer Engineering}, \orgname{The University of British Columbia}, \orgaddress{\street{2332 Main Mall}, \city{Vancouver}, \postcode{V6T 1Z4}, \state{British Columbia}, \country{Canada}}}


\title{An experience-based classification of quantum bugs in quantum software}

\date{September 2024}

\abstract{
As quantum computers continue to improve in quality and scale, there is a growing need for accessible software frameworks for programming them. However, the unique behavior of quantum systems means specialized approaches, beyond traditional software development, are required. This is particularly true for debugging due to \emph{quantum bugs}, i.e., bugs that occur precisely \emph{because} an algorithm is a quantum algorithm. Pinpointing a quantum bug's root cause often requires significant developer time, as there is little established guidance for quantum debugging techniques. Developing such guidance is the main challenge we sought to address. In this work, we describe a set of 14 quantum bugs, sourced primarily from our experience as quantum software developers, and supplemented by analysis of open-source GitHub repositories. We detail their context, symptoms, and the  techniques applied to identify and fix them.  While classifying these bugs based on existing schemes, we observed that most emerged due to unique interactions between multiple aspects of an algorithm or workflow. In other words, they occurred because more than one thing went wrong, which provided important insight into why quantum debugging is more challenging. Furthermore, based on this clustering, we found that---unexpectedly---there is no clear relationship between debugging strategies and bug classes. Further research is needed to develop effective and systematic quantum debugging strategies.}

\maketitle

\section{Introduction}

Quantum software engineering is an emerging field defined by the use of ``sound engineering principles for the development, operation, and maintenance of quantum software" \cite{zhao2021horizons} and associated libraries and documentation \cite{zhao2021horizons,bisicchia2024quantummechanicsquantumsoftware}. While it bears many similarities to classical software engineering, the unique operation of quantum computers and the structure of quantum algorithms necessitates special considerations. Debugging, an essential practice across software engineering, is a major part of the workflow where this applies, as quantum software developers experience challenges above and beyond the classical bugs that arise: quantum bugs.

We adopt the definition of quantum bugs as bugs that occur precisely \emph{because} an algorithm is a quantum algorithm. As a consequence, they require domain-specific knowledge to fix \cite{odmshortpaper}. 
Bugs caused by incorrect usage of gates and measurements in the low-level circuit model, coupled with the frequent need to combine frameworks with different conventions, can lead to unexpected behavior that is challenging to diagnose. Quantum algorithms rely on superposition and entanglement, and our limited intuition for these features means certain bugs fly under the radar until a function is applied in a context that reveals them. Multiple studies found that a common symptom of quantum bugs is \emph{incorrect program output} \cite{zhao2021bugs4q,paltenghi2022bugs}, implying one must know the expected result before even considering the possibility of a bug. Detecting bugs in a hardware setting is further complicated by the presence of noise that may exacerbate (or worse, conceal) underlying programming errors\footnote{In this work, we will not discuss debugging issues due to hardware noise.}.

The aforementioned issues motivate the development of dedicated tools and targeted guidance for debugging quantum programs. At present, a concrete workflow for pinpointing the root cause of quantum bugs is lacking, and the debugging process is guided largely by experience. We sought to use our own experience to address this by analyzing a variety of quantum bugs and their fixes.

This work begins with an overview of existing studies and classification systems of bugs in quantum programs. Building on an earlier position paper \cite{odmshortpaper}, we then describe a set of 14 quantum (or quantum-related) bugs. Most are bugs we experienced firsthand while implementing quantum algorithms. We supplement this with a selection of quantum bugs from GitHub that fit our definition. We detail the bugs' context, symptoms, and the debugging strategies applied to fix them.

A major consequence of our analysis is that many quantum bugs sit at the intersections of categories in the existing taxonomies; in some cases, they did not fit within any category, which emphasizes the increased complexity of quantum bugs. We propose an alternative bug classification scheme, based on these observations, that emphasizes the intersection of root causes.

Our analysis also provided insight into which debugging strategies were most effective for different types of bugs. Notably, we found there is no single best method for all bugs, even within the same category, and that the effectiveness of the investigated methods varied significantly. Nevertheless, we organize the methods into a flowchart as a starting point for developers who suspect they are experiencing a quantum bug.

Overall, our work highlights that quantum debugging remains in its early stages and that there is a need for dedicated debugging tools, in particular targeting bugs that arise at the intersection of different categories. 
By that, we hope to inspire others to contribute to the development of such tools.

\section{Background}

Before proceeding, we recall our definition of quantum bugs to make the important distinction between  (a) bugs in the implementation or usage of a quantum programming framework, and (b) bugs that occur in the implementation of quantum algorithms within these frameworks. 
The former we term \emph{quantum-related bugs}. Such bugs may be associated with the underlying quantum aspects of the program, but do not necessarily require domain-specific knowledge to solve (for instance, errors in control flow or misuse of library functions can happen in the context of classical programming).

Debugging methods for quantum computing software necessarily evolve alongside quantum software development. A growing body of research has emerged over the years, leading to multiple classification schemes of quantum and quantum-related bugs, as well as techniques to detect them.
Both are reviewed in this section.

\subsection{Classification schemes}
To debug issues that are specific to quantum programming, a good understanding of their causes and underlying reasons is crucial.
Such research is facilitated by having access to a diverse set of real bugs (and their fixes) encountered by developers. Concrete examples provide intuition into common patterns, enabling us to augment our tools with features that help avoid them. 

To that end, different \emph{databases} to collect representative sets of bugs based on open-source quantum software repositories were proposed~\cite{campos2021qbugs, zhao2021bugs4q}.
These databases help shed light on prominent and frequent issues developers have.   Finding the underlying common patterns in such databases was tackled by later works, which sought to group related bugs together. Through these works, two different types of classification schemes emerged.
The first type of classifies bugs based on their \emph{root cause}. For example,  categories such as ``API-related", ``incorrect application logic", ``math-related", and ``others" are used in Refs~\cite{paltenghi2022bugs, luo2022comprehensive}.

The second type groups bugs based on \emph{where} in the quantum algorithm the actual problems lie.
Within these schemes, the granularity varies significantly among the proposed approaches:
 Refs.~\cite{huang2019statistical, huang2019towardcorrect} group by ``inputs", ``operations", and ``outputs" as their top-level categories; similarly, Ref. \cite{zhao2021identifying} groups by ``initialization", ``gate operation", ``measurement", and ``deallocation".
A slightly more fine-grained classification that also covers the quantum-classical interaction is proposed in Ref.~
\cite{paltenghi24_survey_testin_analy_quant_softw}. There, ``gate errors", ``measurement issues", ``miscellaneous", ``circuit size", ``quantum-classical interface", and ``qubit initialization and layout" are top-level categories.
Lastly, Ref.~\cite{aoun2022bugcharacteristicsquantumsoftware} proposes a classification that goes one step further by considering aspects such as visualization and monitoring. 
Ref.~\cite{aoun2022bugcharacteristicsquantumsoftware} does not follow a hierarchical categorization and uses the following categories: ``compiler", ``gate operation", ``simulator", ``state preparation", ``measurement", ``post-processing", ``pre-processing", ``pulse control", ``error mitigation", ``quantum cloud access", ``visualization", ``monitoring", ``other".

However, these prior studies only tell part of the story:
\begin{itemize}
    \item As the databases comprised mainly bugs reported to frameworks, most pertain to bugs \emph{within} the framework, rather than bugs that researchers and practitioners experience when using them to \emph{develop new algorithms}.
    \item Most bugs stem from one programming framework (Qiskit \cite{Qiskit}), 
    limiting the scope to primarily a single user base and programming style (object-oriented gate-model algorithms). As such, many bugs are specific to how frameworks manage resources (e.g., the need to manage classical bits that store measurement outcomes causes bugs related to classical register management in Qiskit code).
\item Lastly, none of the existing categorizations accounts for bugs that occur at the intersection between different aspects of a quantum program. 
By focusing on single categories of \emph{where} a bug happens or what its underlying \emph{root cause} is, the identification of more complex patterns among multiple bugs is hindered. 

\end{itemize}
 
\subsection{Debugging methods and tools}\label{sec:debugging_methods}

The knowledge gained from deriving classification schemes can be used to develop better methods and tools to detect and fix bugs---and, ideally, make the life of future (quantum) software developers easier.
Of course, the expertise, methods, and tools from decades of classical software debugging, such as backtracking, unit testing, and runtime data analysis are widely used in quantum computing \cite{garcia2023quantum, long2023testing, miranskyy2020bugfree, miranskyy2021testing, zhao2021horizons, paltenghi24_survey_testin_analy_quant_softw}. These practices are especially natural since quantum programs are often implemented within the same classical software development tools and programming languages.
Therefore, this section explicitly reviews \emph{quantum-specific} debugging methods and tools.
To make their debugging capabilities more accessible, a running example is introduced next.\\

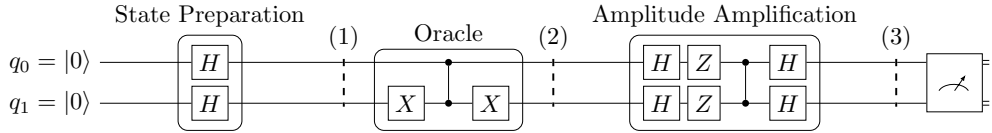
\begin{figure}[t]
    \centering
   \resizebox{\linewidth}{!}{
      \begin{tikzpicture}
                \begin{yquant*}

                    qubit {$q_0=\ket{0}$} q;
                    qubit {$q_1=\ket{0}$} q[+1];

                    [this subcircuit box style={rounded corners, inner ysep=4pt,  label={State Preparation}}]
                    subcircuit {
                        qubit {} q[2];
                        box {$H$} q;
                    } (q[0-1]);
                    [thick, label=(1), dashed]
                    barrier (-);

                    [this subcircuit box style={rounded corners, inner ysep=4pt,  label=Oracle}]
                    subcircuit {
                        qubit {} q[2];
                        x q[1];
                        zz (q[0-1]);
                        x q[1];
                    } (q[0-1]);
                    [thick, label=(2), dashed]
                    barrier (-);

                    [this subcircuit box style={rounded corners, inner ysep=4pt,  label={Amplitude Amplification}}]
                    subcircuit {
                        qubit {} q[2];
                        box {$H$} q[0];
                        box {$H$} q[1];
                        box {$Z$} q;
                        zz (q[0-1]);
                        box {$H$} q;
                    } (q[0-1]);
                    [thick, label=(3), dashed]
                    barrier (-);
                    
                    measure (q[0-1]);

                \end{yquant*}
            \end{tikzpicture}}
    \caption{Grover's algorithm with two qubits.}
    \label{fig:grover_circuit}
\end{figure}

\begin{example}\label{ex:grover}
Suppose we wish to use Grover's algorithm~\cite{grover1996} to identify the marked 2-qubit basis state $\ket{01}$ with high probability.
In general, Grover's algorithm consists of three elements: a state preparation step, an oracle that marks the basis state representing the marked element with a negative phase, and the amplitude amplification step. The full algorithm is visualized in \autoref{fig:grover_circuit}.
\end{example}

\subsubsection{Assertions in quantum software}

A common approach in classical debugging and testing is the use of breakpoints and assertions, and these are valuable in quantum programs as well.
Assertions allow developers to validate a property of a program or variable's state.
Of course, when executing a quantum algorithm on hardware, one can't simply stop the computation and inspect the state\footnote{unless one has the patience and resources to regularly perform full tomography.}. 
Even on a simulator with access to the state vector or density matrix, the results can be challenging to parse, except for small cases. This motivates providing built-in tools with outputs that are simple to interpret. 

Assertion-based quantum debugging was first explored in Ref.~\cite{huang2019statistical}, which proposes a framework implemented in Scaffold \cite{scaffold} based on statistical tests. The choice of assertions is motivated by case studies and a taxonomy of six types of bugs and their suggested defense mechanisms: incorrect initial quantum values, operations, composition of operations through iteration or recursion, mirroring (uncomputation), and incorrect classical input parameters. 

To mitigate these bugs, Ref.~\cite{huang2019statistical} provides three assertions to test properties of a quantum register: \texttt{assert\_classical}, \texttt{assert\_superposition}, and \texttt{assert\_entangled}. Depending on the suspected nature of a bug, assertions can be invoked as pre-conditions or post-conditions, just like classical assertions.

A statistical approach using a $\chi^2$ test is used for validation. This requires halting the program mid-execution for measurement, which can add significant overhead. 

In an ideal world, we could copy a state and test it separately, but this is prohibited by the no-cloning theorem. One workaround explored in Ref.~\cite{wang2023debugcloning} is \emph{approximate cloning}.
Other work incorporates projections to auxiliary qubits that are measured to validate assertions without terminating the program~\cite{li2020proq, liu2020dynamic}. This enables testing of multiple assertions in the same program run, but assumes auxiliary qubits are available (and that the additional operations do not introduce more bugs).

A major challenge with assertion-based quantum debugging is that a programmer must strategically place assertions where they are likely to reveal the source of a bug.
This was addressed in recent work that extended a novel assertion-based framework~\cite{rovara2024frameworkdebuggingquantumprograms} with methods for automatically refining assertions, by either moving them or placing new assertions in other locations~\cite{rovara2024automaticallyrefiningassertionsefficient}.
Using classical simulation, these assertions are evaluated, and automated analysis methods are used to diagnose the underlying root cause to guide users to their bugs.\\

\begin{example}
Different types of assertions could be used to verify the algorithm described in \autoref{ex:grover}.
At step (1) in \autoref{fig:grover_circuit}, an \texttt{assert\_superposition} can validate that all qubits are in a uniform superposition (in this case, it is true).
Next, \texttt{assert\_entanglement} can be used at step (2). Here, we presume the oracle introduced some amount of entanglement; otherwise, the oracle creation went wrong, or no basis state is marked at all.
On the other hand, \texttt{assert\_entanglement} must evaluate to false at step (3), since for the 2-qubit case, only one Grover iteration is required (for more qubits, this is unlikely, and we would expect the assertion at (3) to also be true, assuming the one at (2) holds).
\end{example}

\subsubsection{Visual inspection and interactive tools}
\label{subsubsection:visualization}

In addition to statistical and projection-based approaches, numerous visual approaches have found their place in the realm of debugging quantum software.
These approaches can be divided into visual inspection of
\begin{enumerate}
    \item the underlying quantum circuit,
    \item the quantum state throughout the circuit, and
    \item the measurement outcome.
\end{enumerate}
\emph{Visual inspection of the quantum circuit} is a simple, yet effective debugging method.
It helps to quickly grasp what quantum gates are applied to which qubits. Most quantum SDKs provide functions to easily print a given quantum circuit.\\

\begin{example}
Suppose we forgot to add one of the Hadamard gates in the \enquote{Amplitude Amplification} block of the quantum circuit shown in \autoref{fig:grover_circuit}, leading to a buggy circuit. Upon visual inspection, a user familiar with Grover's algorithm would realize the amplitude amplification block ``looks wrong", and easily identify which gate is missing.\\
\end{example}

On the contrary, \emph{visual inspection of the quantum state} requires a bit more effort.
To generate the quantum state throughout a given quantum circuit, a  quantum simulator (such as, e.g.,~\cite{zulehner2018advancedsim, viamontesImprovingGatelevelSimulation2003} or the ones provided by~\cite{Qiskit, pennylane}) must be used---again, usually provided by most quantum SDKs.\\

\begin{figure}
    \begin{subfigure}[t]{0.3\textwidth}
    \resizebox{\linewidth}{!}{
    \begin{tikzpicture}
 \begin{axis}[
        ymin=-0.6, ymax=0.6,
        ybar,
        symbolic x coords={00, 01, 10, 11},
        xtick=data,
        ylabel={Amplitude},
        xlabel={State},
        axis lines=left,
        ymajorgrids=true,
        grid style=dashed,
        bar width=20pt,         
        enlargelimits=0.1,     
        font=\LARGE,           
        tick label style={font=\LARGE},  
        label style={font=\LARGE}        
    ]
        \addplot coordinates {(00,0.5) (01,0.5) (10, 0.5) (11,0.5)};
    \end{axis}
    \hfill
    \end{tikzpicture}}
    \caption{State at step (1).}
    \label{fig:state_step1}
    \end{subfigure}
    \hfill
        \begin{subfigure}[t]{0.3\textwidth}
    \resizebox{\linewidth}{!}{
    \begin{tikzpicture}
 \begin{axis}[
        ymin=-0.6, ymax=0.6,
        ybar,
        symbolic x coords={00, 01, 10, 11},
        xtick=data,
        ylabel={Amplitude},
        xlabel={State},
        axis lines=left,
        ymajorgrids=true,
        grid style=dashed,
        bar width=20pt,         
        enlargelimits=0.1,     
        font=\LARGE,           
        tick label style={font=\LARGE},  
        label style={font=\LARGE}        
    ]
        \addplot coordinates {(00,0.5) (01,0.5) (10, -0.5) (11,0.5)};
    \end{axis}
    \hfill
    \end{tikzpicture}}
    \caption{State at step (2).}
    \label{fig:state_step2}
    \end{subfigure}
        \hfill
        \begin{subfigure}[t]{0.3\textwidth}
    \resizebox{\linewidth}{!}{
    \begin{tikzpicture}
 \begin{axis}[
        ymin=0, ymax=1.0,
        ybar,
        symbolic x coords={00, 01, 10, 11},
        xtick=data,
        ylabel={Relative Frequency},
        xlabel={State},
        axis lines=left,
        ymajorgrids=true,
        grid style=dashed,
        bar width=20pt,         
        enlargelimits=0.1,     
        font=\LARGE,           
        tick label style={font=\LARGE},  
        label style={font=\LARGE}        
    ]
        \addplot coordinates {(00,0.0) (01,0.0) (10, 1.0) (11,0)};
    \end{axis}
    \hfill
    \end{tikzpicture}}
    \caption{Measurement outcome.}
    \label{fig:meas_outcome}
    \end{subfigure}
    \caption{Visual inspection of the quantum states at steps (1)/(2) and the measurement outcome of \autoref{fig:grover_circuit}.}
    \label{fig:grover_states}
\end{figure}
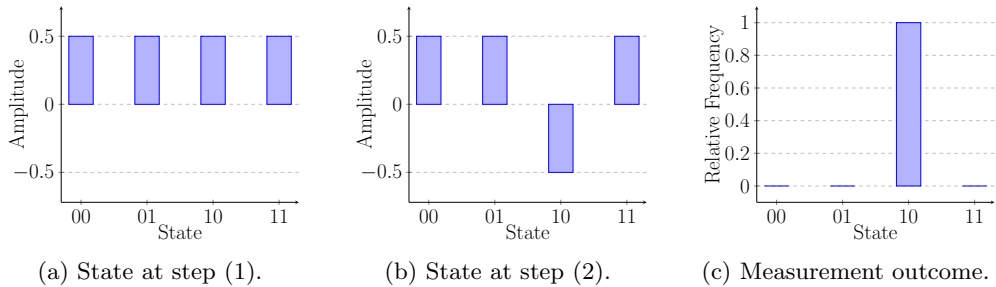

\begin{example}\label{ex:grover_vis_state}
The quantum state's evolution throughout a circuit can be tracked and visualized. 
In \autoref{fig:grover_states}, the quantum states after steps (1) and (2) of \autoref{fig:grover_circuit} are shown. 
Here, \autoref{fig:state_step1} shows the quantum state after the superposition:  all basis states have the same amplitude of $0.5$.
However, \autoref{fig:state_step2} actually reveals a bug: the amplitude of basis state $\ket{10}$ is marked with a phase, while the actual marked element was $\ket{01}$ as described in \autoref{ex:grover}.
Apparently, a \emph{qubit ordering} bug has happened and the indices were mixed up, which is a typical error in quantum software.\\
\end{example}

Lastly, \emph{visual inspection of the measurement outcome} is an often-used debugging technique.
Although it does not provide more information than the counts returned from shot-based execution on a quantum simulator or device, the graphical aspect often helps developers interpret the outcome.\\

\begin{example}
The measurement outcome of the Grover circuit example is shown in \autoref{fig:meas_outcome}.
Due to the amplitude amplification, the probability of measuring the marked element becomes $100\%$ while all other basis states have a measurement probability of $0\%$.
We note this method also reveals the qubit order error, similar to \autoref{ex:grover_vis_state}.\\
\end{example}

There also exist specialized visualization tools to make these approaches  more accessible.
One example is Quirk \cite{quirk}, a web-based graphical circuit composer with numerous options for viewing real-time diagnostic information in a circuit, such as amplitudes, measurement outcome probabilities, and Bloch spheres.  
Quirk can help sanity check small quantum algorithms; however it does not support code upload and is limited to 16 qubits. Furthermore, the amount of visual information can become unmanageable well before then.
CircInspect~\cite{khan2024circinspect} is a web-based graphical circuit composer that proposes a more scalable approach for larger circuits. 
It performs automated abstraction of quantum subroutines (written in an in-browser code editor with PennyLane \cite{pennylane}), so a user can isolate and zoom into different parts of a circuit. The circuit can be updated in real-time or, similar to a traditional debugger, it also allows users to set breakpoints and step through the code. However, it currently only supports PennyLane. The Classiq \cite{classiq} platform has similar function abstraction capabilities, though it does not currently have debugging support.

On a larger scale, Refs.~\cite{metwalli2022debugging,metwalli2023cirquo} introduced Cirquo, a Qiskit tool designed to slice circuits vertically and horizontally into subcircuits which are verified individually. Cirquo, which aims to provide a more systematic means of debugging, can be invoked by adding instructions in a quantum circuit. The developers of Quantivine \cite{wen23_quant}, a tool for visualizing large-scale circuits, explain how it can be used to drill down into subroutines and identify the location of a bug. Basic debugging tools in Visual Studio can be invoked when running Q\#, which also contains special machinery for dumping machine state, assertions, and developing unit tests (which is necessary as Q\# functions themselves do not have any side effects). Finally, a visualization tool called the Qiskit Trebugger \cite{trebugger} (transpiler debugger) was developed for inspecting transpiler passes applied during circuit optimization. We note that, as with the empirical studies, many tools are tied to a specific framework, though the strategies they use could in principle be applied more broadly.

\subsubsection{Equivalence checking}
In the following, we provide a high-level overview of concepts from formal methods and program verification.
A full discussion is beyond the scope of this article, and we refer the interested reader to recent review articles on the subject \cite{ying2019toward, chareton2022formal, lewis2023formal}.
Verification is the process of demonstrating, through analytical proof or a probabilistic procedure, that a quantum program works as intended. 
Ref.~\cite{chareton2022formal} distinguishes between \emph{low-level verification}, at the level of compilers and equivalence checking (EC), versus \emph{high-level verification} which determines whether an algorithm itself is correct. 

The most straightforward means to perform EC of two circuits is to compare their matrix representations. The exponential size, coupled by the fact that the problem of identity checking is QMA-complete \cite{janzing2003identity}, makes scalable EC a challenge. Different approaches, such as path sums \cite{amy2019towards}, decision diagrams \cite{burgholzer2021qcec, hong2022equivalence, peham2022equivalenceparadigms}, the ZX-calculus \cite{peham2022equivalence, peham2023equivalence}, and simulative techniques \cite{burgholzer2020random} have all been applied to EC.
EC is particularly important in compiler workflows, where many optimization passes are sequentially applied to a quantum circuit. VOQC \cite{hietala2021voqc}, built on SQIR (discussed below), was used to implement verified circuit optimization and qubit mapping protocols. In a similar vein, CertiQ \cite{shi2020certiq} was developed for automated verification of Qiskit transpiler passes.\\

\begin{example}
    Returning to our Grover example in \autoref{fig:grover_circuit},
     suppose we wish to run the circuit on a quantum device with a native gateset of \{$H$, $Z$, $SX$, $R_Z$, and $CZ$\}.
    The circuit must be compiled such that each gate which is not part of the native gateset is replaced by a functionally equal sequence of native gates. 
    In this case, only the oracle and, more specifically, its $X$ gates are not native---however, they can be easily substituted with two $SX$ gates since $SX\cdot SX\equiv X$.
    EC can quickly verify if the compilation procedure conserved the functional equivalence and ensure no bugs occurred, such as, e.g., each $X$ gate being mistakenly substituted by just one $SX$ gate.\\
\end{example}

\subsubsection{Formal verification}
While EC answers the question  ``are two circuits functionally equivalent?", it does not address the question ``does this circuit have the expected behaviour?". 
Formal methods for verification address the latter.
The core of many verification procedures involves the manipulation of mathematical models of a quantum computation, usually done by an automated theorem prover. The quantum-specific parts are often embedded languages on top of existing classical frameworks in order to leverage built-in functionality such as symbolic matrix manipulation. 
Beneath the embedded language is the \emph{semantics} describing how the computation is managed mathematically. The idea is then to prove that a program written by a user implements the correct mathematical operation. 

Formal verification generally requires a user to work through their program by directing a proof assistant with formal mathematical statements. Each function or gate in a circuit has an associated mathematical expression, and one can verify, much as one would by hand, the correctness of the composition of operations. This is the approach taken by frameworks such as SQIR \cite{hietala2021proving} and QWIRE \cite{qwire} (which are built on the classical theorem prover Coq). Quantum Hoare Logic, based on the idea of quantum while programs, validates pre- and post-conditions of a program's operation using a density matrix formalism (for a thorough description, see Ref.~\cite{ying2019toward}).
Other procedures use different mathematical representations. For example, Refs.~\cite{qbricks} and \cite{amy2019formalphd} are based on the path-sum formalism, wherein computations are expressed using polynomials that represent how they act on computational basis states.

Given the variety of formal verification methods available, the reader may wonder why debugging tools are necessary. A major shortcoming of formal verification is that it has been primarily limited to proving the correctness of \emph{textbook} algorithms, i.e., algorithms for which we can write down a solution in a closed form to compare against. But even for well-known algorithms such Shor's algorithm, it was only in 2022 that a full, formally verified implementation was developed~\cite{peng2022formally}. It is less clear how to  apply these tools to novel algorithms. There is also a great need for more automation and integration to make such methods accessible in common quantum programming frameworks.\\ 

\begin{example}
Returning to our Grover example from \autoref{fig:grover_circuit}, suppose we wish to use formal verification to validate its output. Subroutines such as the Hadamard transform and amplitude amplification can be verified independently. 
Therefore, the focus should lie on validating the oracle implementation to ensure it implements the intended functionality---in this case, marking the  $\ket{01}$ state with a phase.
To this end, formal verification tools could be used to prove a mathematical condition such as $O(\ket{01}) = -\ket{01}$.
In this case, the procedure would fail, since, as shown in the previous examples, the qubit order is mixed up.\\
\end{example}

\section{Case studies}
\label{sec:case-studies}

The focus of this work is analyzing bugs that occur when implementing a quantum algorithm, independently of the framework used (to the extent this is possible). 
In this section, we describe two endeavours which were the source of numerous first-hand bugs. These are supplemented with a selection of bugs from GitHub issues of open-source repositories that relate to bugs in algorithm implementations. Throughout, we give names to specific bugs using a capitalized identification string (e.g., QORDER represents  qubit-ordering bugs).

\subsection{Case study 1: Shor's algorithm}

The context of this case study is developing an implementation of Shor's algorithm using Catalyst \cite{catalyst}, a library enabling just-in-compilation of PennyLane \cite{pennylane} programs. 

The quantum subroutine at the core of Shor's algorithm is quantum phase estimation (QPE) \cite{shor1997shorsalgo}. QPE contains an inverse quantum Fourier transform (QFT), and in Shor's algorithm, uses circuits for modular arithmetic. 
In QPE, the number of qubits in the register the QFT is applied to (the \emph{estimation register}) affects the quality of the results, so there is a tradeoff between precision and resources. However, there is a well-known implementation  requiring only one estimation qubit \cite{parker2000efficient}, assuming one can perform mid-circuit measurement and reset, and apply operations with parameters controlled on classical values. This feature is available from several hardware providers \cite{corcoles2021exploiting, moses2023race}, and was recently used to perform phase estimation in practice \cite{lubinski2022advancing}.

Even during purely classical simulation, measure-and-reset QPE can save significant resources. This project used circuits from Ref.~\cite{beauregard2003circuit}, where the measure-and-reset QPE can be found in its Figure 8.
In this circuit, measurement outcomes are used to estimate a phase, which is then converted into a guess for the \emph{order}. 
Measurement samples are obtained after applying a rotation gate whose angle depends on previous measurement outcomes via a gate $\mathbf{R}_k$. However, Ref.~\cite{beauregard2003circuit} does not explicitly define $\mathbf{R}_k$, so the author referred to earlier work (Ref.~\cite{parker2000efficient}), which contains an alternative version of the circuit. The expression for $\mathbf{R}_k$ was in the caption of Fig. 2 \cite{parker2000efficient}.

However, there are numerous differences in the two circuits, the most notable being which measurement outcome index corresponded to which controlled $U_a^{2^k}$ operation. While implementing this subroutine, one must take care to ensure the qubits and measurements are processed in the correct order, and those measurement outcomes are used in the correct order to compute the estimated phase. The discrepancies in conventions inevitably led to an initially-incorrect algorithm output.

We expect most, if not all, quantum software developers have run into similar bugs related to \emph{qubit ordering} (QORDER). While differing conventions across frameworks are a major source, it can also happen within a single framework (e.g., not realizing that a QFT implementation has SWAPs at the end (QFT); or encoding an integer value into a qubit register with the opposite endianness that a subroutine expects).
This instance of QORDER was debugged with unit testing, augmented with visualization of state and measurement outcomes. QORDER also occurred elsewhere in the Shor implementation, such as bit ordering in the modular exponentiation routines (also solved by unit testing, and inspection of measurement outcomes to ensure arithmetic was correct and the  measurements were processed with the correct endianness). 

A variety of other bugs arose:
\begin{itemize}
    \item (RESET) The measured qubit was not reset to $\ket{0}$ before being reinitialized and applying a Hadamard. The circuit in Ref.~\cite{beauregard2003circuit} shows reset explicitly, but that in Ref.~\cite{parker2000efficient} does not; and PennyLane's \texttt{measure} function defaults to \texttt{reset=False}.
    \item (XBASIS) The wrong measurement basis was used. Ref.~\cite{beauregard2003circuit} suggests a measurement in computational basis, whereas Ref.~\cite{parker2000efficient} places a Hadamard gate immediately before the measurement.
    \item (UNCOMP, XGATE) Incorrect gates led to auxiliary qubits not being properly uncomputed in modular exponentiation routines; these qubits were later reused.
\end{itemize}

UNCOMP was identified through unit testing and visualization of measurement samples, though we note it is an excellent use case for assertions that validate a register of qubits is all $\ket{0}$s. XBASIS and XGATE were solved here by visualization. These issues are common, however, and EC and formal verification could also be used in most cases. RESET was diagnosed by visualizing the circuit and the measurement outcomes (and as mentioned earlier, EC could also be used). 

While one can argue that the developer should simply be more careful, such bugs expose a number of issues. One is that inconsistent conventions across frameworks and resources consumes significant developer time (we discuss this further in \autoref{sec:classification}).
Another is that algorithms with dynamic circuits would benefit from dedicated tools. For instance, a graphical interface with which a user can view a circuit and interactively select from a drop-down list of measurement options. These choices could then update other parts of the diagram. However, even with such a tool errors can go undetected.

EC for dynamic circuits is a viable solution~\cite{hong2022equivalence, burgholzer2022nonunitaries} (though such routines are themselves not immune to bugs; see PERM below). A simple option is to add qubits and invoke the deferred measurement principle to construct equivalent non-dynamic circuits, though this increases the number of qubits. As noted explicitly in Ref.~\cite{burgholzer2022nonunitaries},  implementing dynamic circuit EC in commonly-used frameworks is still outstanding. Moreover, the burden of implementing both versions of the circuit remains with the same user, i.e., the person already reconciling different conventions that led to the bug.

\subsection{Case study 2: implementing a variational eigensolver on hardware}

The reader is referred to the more detailed bug descriptions in Ref.~\cite{odmshortpaper}. Here, we provide only a brief explanation of the bugs and the debugging strategies employed.

\begin{itemize}
    \item (DECOMP) A custom decomposition of an operation contains incorrect gates.
    \item (GPHASE) A gate, implemented up to a global phase, is applied controlled on  another qubit. The global phase becomes a local phase, producing incorrect output.
    \item (VRZ) Circuits were passed from one framework to another to leverage its transpiler. The transpiler removed $R_Z$ gates at the end of the circuit. The transpiled circuit was used for measuring the expectation value of a Hamiltonian that required additional basis measurements; the measurement results were incorrect.
    \item (VQEM) A hardware provider's transpiler optimized away the gates that were added for error mitigation routines. This is a unique instance in which the bug symptom was \emph{correct output}.
\end{itemize}

An interesting feature shared by many of these bugs is that the symptoms only manifested when something happened in more than one ``where". For example, the removal of $R_Z$ is only problematic when the measurement basis changes. Similarly, the GPHASE bug within the algorithm can only be detected when measuring in a non-computational basis. DECOMP, VRZ, and VQEM are all related to transpilation, in addition to the algorithm and measurement.

The debugging approaches were as follows: 

\begin{itemize}
    \item (DECOMP) Visualization (circuits); EC
    \item (GPHASE) EC 
    \item (VRZ) Visualization (circuits); EC
    \item (VQEM) Visualization (program)
\end{itemize}

In analyzing these two case studies, EC emerges as a go-to method for debugging quantum programs. EC is straightforward for frameworks to incorporate, in particular in a transpiler. However, it relies on a correct reference implementation. If there is a bug in the implementation of a gate, and that gate is used in two different (but equivalent) circuit decompositions, equivalence is preserved, but the underlying implementation is still wrong. Moreover, there are exceptions: EC does not detect the VQEM bug because the circuits before and after error mitigation are equivalent by design. VQEM was rectified by visual inspection (of the circuit returned by the hardware provider, and the provider's documentation).

\subsection{Other bugs}

This section showcases a selection of quantum bugs (and some common quantum-related bugs) reported in GitHub issues. They were identified by searching the repositories of the following open-source frameworks (filtering by ``bug" when the option was available):
\begin{itemize}
    \item 3 general-purpose frameworks (Qiskit~\cite{Qiskit}, PennyLane~\cite{pennylane}, Cirq~\cite{cirq})
    \item 3 transpilers (TKET~\cite{tket}, staq~\cite{staq}, BQSKit~\cite{bqskit})
    \item 3 other libraries (stim~\cite{gidney2021stim}, mitiq~\cite{mitiq}, and a subset of Munich Quantum Toolkit~\cite{willeMQTHandbookSummary2024} repositories) 
\end{itemize}

These frameworks were chosen based on their prevalence, our research areas, and in some cases, our experience as contributors. While most issues are bug reports for quantum-related bugs in library code, we identified a few unique bugs, as well as numerous variants of our own bugs.

\subsubsection{(SP) PennyLane Issue \#5099}

The adjoint of a state preparation routine, \texttt{qml.StatePrep}, was applied controlled on another qubit \cite{pennylane5099}. It was expected that if \texttt{qml.StatePrep} creates state $\ket{\psi}$, then its adjoint applied to $\ket{\psi}$ should result in $\ket{0}$. However, an additional global phase was introduced, causing GPHASE to manifest when a control qubit was involved.

There are multiple facets to this bug but the core, which we denote by SP, is that state preparation (for a user-specified input state) was used as an arbitrary quantum operation. However, \texttt{qml.StatePrep} was designed for use only at the beginning of a circuit. State preparation routines will send $\ket{0}$ to a desired state $\ket{\psi}$, but the action on other basis states may be arbitrary. This was further complicated by the implementation details of the framework. The mid-circuit instance of \texttt{qml.StatePrep} was decomposed using the routine from \cite{mottonen2004transformationquantumstatesusing}, which at the time was only implemented up to a global phase.

We infer from the report that the issue was detected by visualizing the output state vector; we note that EC up to a global phase would also reveal it. The framework's implementation was later updated to incorporate the correct global phase.

\subsubsection{(MAPINFO) pytket-qiskit Issue \#123}
In this issue \cite{pytket123} a user wanted to map a quantum circuit using a method in TKET~\cite{tket}, then convert it into a \texttt{Qiskit.QuantumCircuit} using the parsing methods provided by TKET's Python package \texttt{pytket-qiskit}.
However, it was noted, that the \texttt{Qiskit.QuantumCircuit.layout} attribute to store mapping information was not filled by the TKET parsing function. 
The result was that the \texttt{Qiskit.QuantumCircuit} did not include layout or mapping information, and the mapping was not applied.

While in some regards this is a quantum-related bug (as it was due to missing classical information), it is another example of a bug that spans multiple categories: transpilation, and cross-framework ``plumbing", since the mapping information was only lost when parsing from TKET to Qiskit. This bug, denoted by MAPINFO, was caught by visual inspection of the \texttt{Qiskit.QuantumCircuit} after parsing.

\subsubsection{(MATRIX) Qiskit Issue \#13118 and MQT QCEC Issue \#10}
Each compiler usually stores the underlying matrix representations of all supported quantum gates.
This represents a single source of truth and a small mistake can lead to various symptoms with no obvious common pattern. 
In the referenced Qiskit issue \cite{qiskit13118}, two of the four parameters  in the matrix representation of the $CU$ gate were accidentally switched; in the MQT issue, a minus sign was missing in the $R_Z$ implementation.

This bug (denoted by MATRIX) is one of the few general quantum-related bugs we discuss.
The Qiskit issue was detected by a user that got unexpected and faulty measurement outcomes when simulating a circuit with heavy transpiler optimization.
Detection of the MQT issue was more subtle since the $R_Z$ information was not much used since
OpenQASM's ``qelib1.inc" translates a $R_Z$ gate to a phase gate, which was correctly implemented, and the tool was mostly used via OpenQASM circuits.

\subsubsection{(PERM) Qiskit Issue \#10457 and MQT QCEC Issues \#251 and \#346}
In many transpilation-related tasks such as quantum circuit mapping and EC, it is necessary to consistently store layout information, i.e., which logical qubit is assigned to which physical qubit.
Layout management is more challenging than it first appears, as it typically changes throughout a circuit's execution.
In the referenced Qiskit issue \cite{qiskit10457}, the mapping information was not correct for a specific mapping routine that tried to find an improved layout for a \texttt{Qiskit.QuantumCircuit} that had a layout previously assigned.
In the first referenced MQT issue \cite{qcec-251} an error was caused when the two to-be-checked quantum circuits had an unequal number of qubits, even though the additional qubits were idle qubits and had no gates applied to them.
In the second \cite{qcec-346}, determining the initial layout of dynamic quantum circuits caused the error, since this is a non-trivial task that is approached by determining the input layout based on a given output layout.

While this bug is primarily related to transpilation, it  also pertains to qubit initialization and measurement, because both were necessarily involved to reveal it.
The Qiskit issue was revealed by visual inspection of the \texttt{Qiskit.QuantumCircuit} and comparing it with the stored \texttt{Qiskit.QuantumCircuit.layout} information which did not match.
Both MQT bugs were revealed by EC; the first one by comparing two circuits where one had an additional idle qubit and the second one by comparing a static and a dynamic circuit that had the same functionality (but EC returned false).

\section{Experience-based quantum bug classification}
\label{sec:classification}

 \autoref{tab:bug_table} provides a summary of the bugs in \autoref{sec:case-studies}, including the bug source and a brief description.
 We found many bugs share common characteristics. Issues with phase arise repeatedly. Qubit management issues (their ordering convention, permuting the order during mapping or after an operation like the QFT) also present frequently.

\begin{table}[htpb]
    \centering
\begin{adjustbox}{angle=90}
    \begin{tabular}{|p{1.5cm}|p{5.6cm}|p{2.4cm}|p{2.1cm}|p{2.8cm}|p{1.2cm}|} 
    \hline
        \textbf{Bug ID} &  \textbf{Brief description} &  \textbf{Source} &  \textbf{Classes} &  \textbf{Symptom} &  \textbf{Strat.} \\ \hline
         GPHASE & A global phase becomes a local phase when an operation is applied controlled on another qubit. & Experience & ALG & Incorrect output & EC, FV \\ \hline
        VRZ & Terminal $RZ$ gates in front of non-$Z$ basis measurements are removed by transpilers. & Experience & ALG, MEAS, TRNS, PLUMB & Incorrect output & EC, FV, VI-P \\ \hline
         VQEM & Gates that are added for error mitigation are removed by transpilers. & Experience & ALG, MEAS, TRNS, PLUMB & Correct output & VI-P \\ \hline
     SP & An adjoint of a state preparation subroutine (implemented up to a global phase) is used mid-circuit and is controlled by another qubit.  & PennyLane GitHub \cite{pennylane5099} & ALG & Incorrect output & VI-S \\ \hline
    RESET & Qubits are not correctly reset after a mid-circuit measurement.  & Experience & INIT, ALG, MEAS & Incorrect output & EC, VI-M, VI-P \\ \hline
    DECOMP & The decomposition of an operation is implemented incorrectly. & Experience & TRANS & Incorrect output & EC, FV \\ \hline
    UNCOMP & Operations on auxiliary qubits are not properly uncomputed. & Experience & ALG & Incorrect output & AS, VI-M, EC, FV \\ \hline 
    QFT & Lack of conventions for the qubit ordering and whether to SWAP after a Quantum Fourier Transform leads to confusion. & Experience, Qiskit Github \cite{qiskit4849} & ALG & Incorrect output & VI-P, EC, FV \\ \hline     
    XGATE & An incorrect gate is applied or its gate direction is wrong. & Experience & ALG & Incorrect output & EC, FV, VI-P \\ \hline XBASIS & An incorrect measurement basis is used. & Experience & ALG, MEAS & Incorrect output & VI-M \\ \hline   
    MAPINFO & Qubit mapping information is not preserved when parsing between frameworks. & TKET-Qiskit GitHub\cite{pytket123} & TRNS, PLUMB & Missing information & AS, EC \\ \hline
    MATRIX & The matrix representation of a gate is incorrect (either due to error, or difference in convention). & Qiskit GitHub \cite{qiskit13118} & GEN & Missing information & VI-S \\ \hline    
    PERM & The permutation of qubits (due to algorithm of mapping) leads to incorrect ordering or interpretation of results. & MQT GitHub \cite{qcec-251} Qiskit GitHub\cite{qiskit10457} & INIT, MEAS, TRNS, PLUMB & Missing information, incorrect output & VI-S \\ \hline
    QORDER & The qubit ordering is incorrect. & Experience & INIT, ALG, MEAS, TRNS, PLUMB & Incorrect output & VI-S \\ \hline     
    \end{tabular}
\end{adjustbox}
    \caption{Selection of quantum bugs. Classes correspond to those in \autoref{tab:our_bug_classes}. Strategies are equivalence checking (EC), code-based assertions (AS), visualization (VI) of measurement outcomes (VI-M), program representation (e.g., circuit or code) (VI-P), or the quantum state (VI-S), and formal verification (FV).}
    \label{tab:bug_table}
\end{table}

 We initially attempted to classify the bugs using the existing taxonomies. However, we observed many of them did not fit well in existing schemes, either because (1) they did not fit into any category, or (2) they fit into multiple categories.  As a concrete example, consider the VRZ bug. The scheme of Ref.~\cite{zhao2021bugs4q} might classify it as either a “gate operation” or “measurement” bug, already highlighting ambiguity. Similarly, Ref.~\cite{paltenghi24_survey_testin_analy_quant_softw} might label it as a “gate error”, “measurement issue”, or place it under “miscellaneous”, while Ref.~\cite{aoun2022bugcharacteristicsquantumsoftware} might categorize it as a “compiler”, “gate operation”, “measurement”, or “other” bug.

To resolve issue (1), we initially proposed a new scheme with six bug categories, which are summarized in \autoref{tab:our_bug_classes}. 
This classification is at first glance similar to the location-based schemes in \cite{huang2019statistical, huang2019towardcorrect, zhao2021identifying,paltenghi24_survey_testin_analy_quant_softw,aoun2022bugcharacteristicsquantumsoftware}. The intention was to augment the location-based approach with additional ``root cause" information. For instance, initialization errors are not limited to the beginning of the circuit; measurement is not limited to the end. TRNS and PLUMB are independent categories to reflect the fact that these processes lead to unique issues.

\begin{table}[]
    \centering
    \begin{tabular}{|c|p{10cm}|}
    \hline
        \textbf{Bug class} &  \textbf{Description} \\ \hline
        INIT & Bugs related to allocation and initialization of qubits, either at the beginning of an algorithm or partway through. \\ \hline
        ALG & Bugs in usage of quantum operations in an algorithm (circuit). \\ \hline
        MEAS & Bugs related to usage or interpretation of measurements, either at the end of an algorithm, or partway through (e.g., mid-circuit measurements). \\ \hline
        TRNS & Bugs that occur either as part of or due to the quantum compilation process, e.g., transpilation. \\ \hline
        PLUMB & ``Quantum plumbing" bugs that arise when writing ``pipes" between different quantum programming frameworks with different conventions. \\ \hline
        GEN & Quantum-related bugs, such as errors in classical control flow, mathematical representation, or framework misuse. \\ \hline
    \end{tabular}
    \caption{Quantum bug classification proposed in this work. Rather than considering these as six distinct classes, we emphasize that many quantum bugs actually fall at the intersection of one or more (see \autoref{fig:venn}). The deliberate consideration of such intersections is thus a distinguishing feature of this scheme.}
    \label{tab:our_bug_classes}
\end{table}

\begin{figure}
    \centering
    \includegraphics[width=0.8\linewidth]{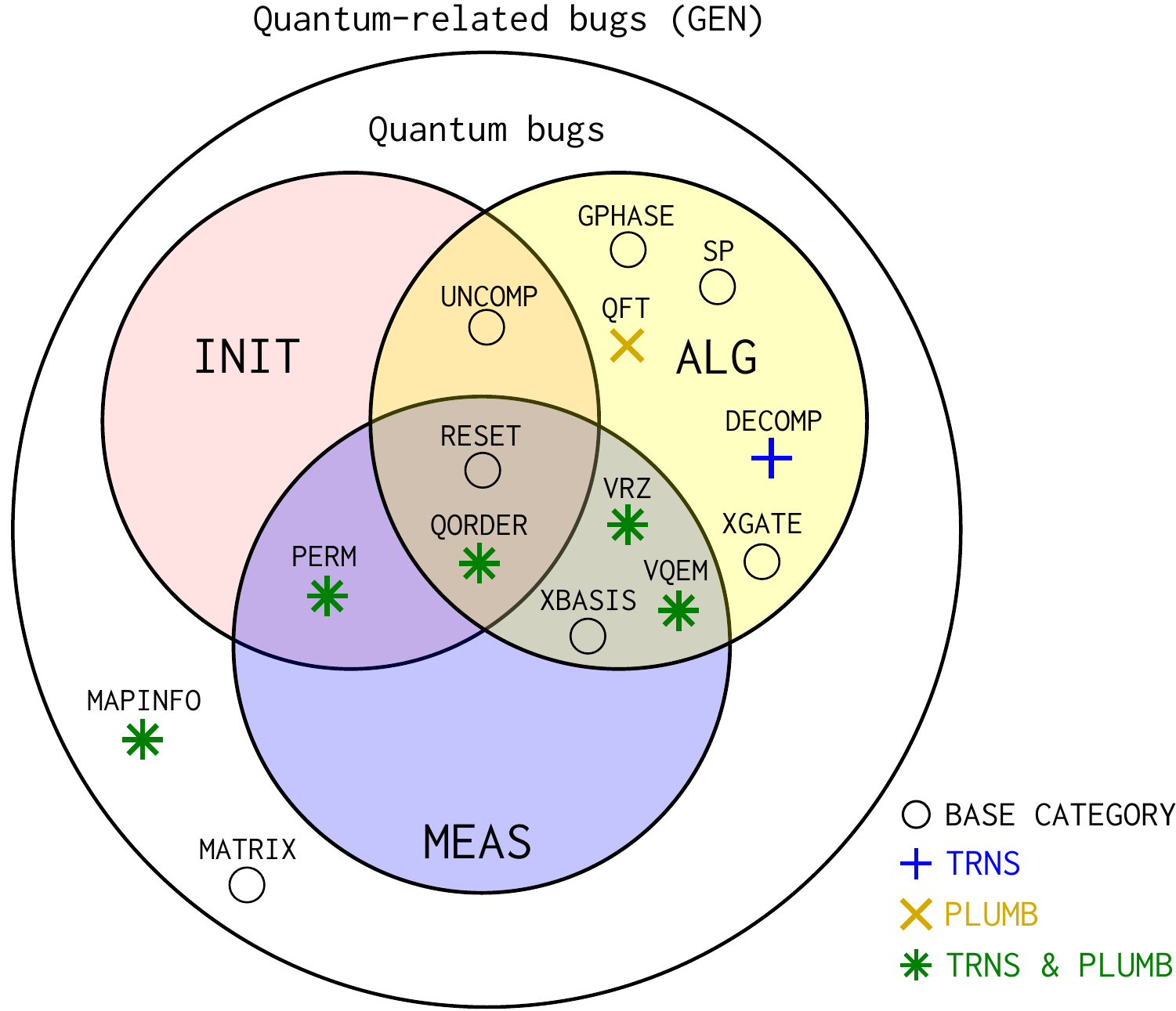}
    \caption{Bugs in \autoref{tab:bug_table} based on the classification in \autoref{tab:our_bug_classes}. Due to dimensionality constraints, \olivia{TRNS and PLUMB are denoted by icon shape} (the stars indicate both). The emptiness of INIT and MEAS does not imply the non-existence of single-category bugs; rather all those we experienced intersected with at least one other category.}
    \label{fig:venn}
\end{figure}

However, upon further analysis, we still found that numerous bugs fell in multiple categories. We thus argue that a hallmark feature of many quantum bugs is that they are ``compound", in the sense that more than one thing must be done incorrectly,  more than one conceptual misconception occurs, or that identifying them requires more than one debugging approach. This makes quantum bugs both harder to detect and harder to solve. To illustrate this, we combine the contents of \autoref{tab:bug_table} and \autoref{tab:our_bug_classes} using a Venn diagram in \autoref{fig:venn}.

From \autoref{fig:venn} we see that all but three quantum bugs span two or more categories: GPHASE, SP, and XGATE (and these three pertain to the most ``quantum" part, i.e., the actual algorithm). Even bugs we initially considered single-class were only triggered or detected due to their interaction with a different class. For instance, RESET initially presents as a purely initialization bug (i.e., after a mid-circuit measurement, a qubit is not correctly re-initialized, and that re-initialization  is the issue). However, we only detect this bug if a qubit is re-used later in an \emph{algorithm} (which may itself have bugs) and \emph{measured} again. Similarly, in VRZ, the removal of RZ gates seems like solely a transpiler-based algorithm issue, but it was only detected when \emph{measuring} in different bases. In this vein, we remark that the absence of points solely in INIT or MEAS does not imply the absence of such bugs, only that we have yet to encounter them.

\section{Implications of the proposed classification}
As visualized in \autoref{fig:venn}, the proposed classification scheme shows clear groups of bugs that fall in the same category or intersections of categories.
However, when comparing their most promising debugging strategies listed in \autoref{tab:bug_table}, there are no clear common strategies.
For example, when considering the three bugs at the intersection of ALG and MEAS (VRZ, VQEM, XBASIS), the suggested debugging strategies are equivalence checking, formal verification, visual inspection of the circuit or program and visual inspection of the measurement outcomes---and, therefore, three out of four of the debugging methods reviewed in \autoref{sec:debugging_methods} are used. The diversity of strategies is even greater for groups that contain more examples.
Consequently, one outcome of this experience-based classification is, that there is no single best debugging method for a given class of bugs.

Coming from a software developer's perspective, this naturally raises the questions ``How should I debug if there is no guidance for what strategy is best in certain cases?".
Although explicit guidance is hard to give, we noticed differences in the frequency in which the debugging methods proved helpful: equivalence checking and visual inspection of the circuit (program) and state are the clear winners in our evaluation. 
However, the reasons for their predominant role might not only be their effectiveness, but also their ease of use: nearly all quantum SDKs provide simple methods to print quantum programs and circuits. Similarly, equivalence checking tools are often provided as Python package that offer native integration with most of the available quantum SDKs, making them highly accessible.

This is unfortunately not true for the remaining techniques, namely assertion in quantum software, the visual inspection of the quantum state, and formal verification. 
All of these methods come with significant overhead.
Assertions have not been integrated well in most prominent quantum SDKs such as Qiskit.
Inspecting quantum states is straightforward in theory, but in practice, it becomes challenging to use simulators to go through quantum programs or circuits in a step-wise process---not even mentioning of the scalability of quantum simulators in general.
Formal verification comes with an even higher effort since this approach has not established itself as a common tool for quantum software developers and, therefore, there is a lack of suitable tooling that integrates well with existing SDKs.

Lastly, the scalability of debugging methods is an important aspect to consider.
When transitioning from the noisy-intermediate scale quantum computers to early fault tolerant computers, quantum programs will grow significantly due to the overhead induced by quantum error correction. 
This naturally raises the question of whether methods and tools that work well for dozens of qubits will still be effective when considering thousands of qubits.
While some techniques may scale, many will quickly become infeasible (such as all three visual inspection approaches).

\begin{figure}[t]
    \centering
    \includegraphics[width=0.35\linewidth]{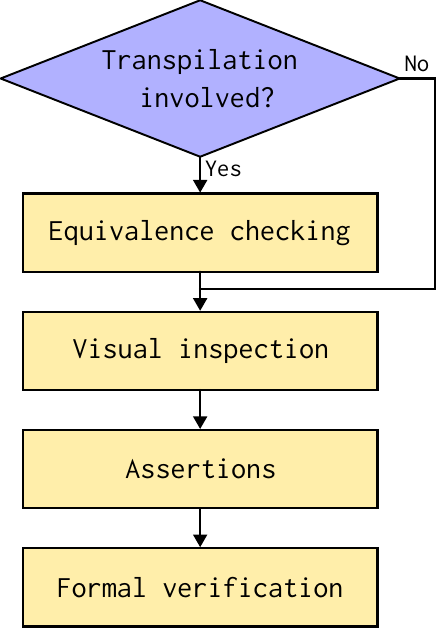}
    \caption{Experience-based flowchart for selecting a debugging strategy, based on effectiveness and usage frequency for the bugs in \autoref{tab:bug_table}.}
    \label{fig:debugging_flowchart}
\end{figure}

Combining these insights and our experience, we recommend the flowchart depicted in \autoref{fig:debugging_flowchart} as a starting point for debugging erroneous quantum program.
When compilation is involved (e.g., two different implementations of a circuit are available), quantum circuit equivalence checking might be a good first choice. 
If this does not reveal any underlying issue, visual inspection is often beneficial.
However, visual inspection may be infeasible for large programs. Assertion testing can then be applied to check against desired criteria at specific parts in a program (e.g., if one has knowledge of properties of the state after a particular subroutine).
If no suspicious-looking part of the circuit can be identified, the last resort would be formal verification.

\color{black}

Of course, all of these outcomes and learnings must be taken with a grain of salt. 
The classification, investigated example bugs, and suggested debugging strategies come with a significant personal and experience-based flavor, and no generality is claimed.
To gain a deeper understanding, more developers should participate in the creation of such classification schemes, using a larger variety of algorithms. 
One option to lessen the effects of personal bias is to perform a large-scale inter-annotator experiment, wherein multiple developers are tasked with classifying the same set of bugs. In addition, such an experiment should include bugs from more open-source repositories and domains within quantum computing software. A more fine-grained set of debugging methods should be included, and additionally, we must also consider how classical debugging methods fit into the picture.

\section{Conclusions}

The \emph{quantum} nature of ``quantum computing'' leads to an entirely different set of ways to make mistakes. 
Debugging quantum programs is challenging, especially as their hallmark symptom, incorrect output, can be hard to interpret.
Our experience leaves us with a few key takeaways:
\begin{enumerate}
    \item Qubit ordering and global phase are common sources of error.
    \item Many quantum bugs sit at the intersection of more than one category. In other words, more than one thing must go wrong.
    \item We found no clear relationship between debugging strategies and bug classes, but equivalence checking and visualization are accessible and work well in most cases.
\end{enumerate}

The example bugs we presented are but the tip of the iceberg. 
Our findings highlight that quantum debugging is still in its early stages, with a particular need for tools that address bugs arising at the intersection of multiple categories.
As the technology becomes more prevalent and the number of quantum algorithms (and languages in which to implement them) grows, comprehensive tools and strategies for debugging quantum algorithms will be critical.

This work can be a starting point for further research and development of dedicated debugging approaches to support quantum software developers in different ways.
Moving forward, \autoref{fig:debugging_flowchart} should be improved and expanded upon to provide more concrete guidance on which strategies to try, depending on the suspected source of the error. This would save significant developer time.
Also, tools for automatically detecting and fixing bugs, such as the recent work of \cite{bugschatgpt}, will likely play an important role in the future (interestingly, the tool's success rate was lowest for ``quantum algorithm bugs"). 
Developing such tools will require surveying a diverse set of developers, to gain insight into how often such bugs occur in practice.

\section*{Acknowledgments}

Both authors thank the students in the Quantum Software and Algorithms Lab at UBC and the Chair for Design Automation at the Technical University of Munich for engaging discussions about programming mistakes, and the anonymous reviewers for their valuable feedback. ODM acknowledges funding from Canada's NSERC, the Canada Research Chairs program, and the Department of Electrical and Computer Engineering at UBC. She is also grateful to Xanadu, where the initial portions of this document were conceived and written, for their hospitality and commitment to well-documented, open-source quantum software. 
N.Q. acknowledges funding from the European Research Council (ERC) under the European Union’s Horizon 2020 research and innovation program (grant agreement No. 101001318), the Munich Quantum Valley, which is supported by the Bavarian state government with funds from the Hightech Agenda Bayern Plus, and has been supported by the BMK, BMDW, the State of Upper Austria in the frame of the COMET program, and the QuantumReady project within Quantum Austria (managed by the FFG).

\bibliography{main}

\end{document}